\documentclass[aip,rsi,preprint,groupedaddress,showmsc,superscriptaddress,floatfix,pra,aps,showpacs,onecolumn,twoside]{revtex4}
\usepackage{amsmath,amsfonts,amssymb,caption,hyperref,color,epsfig,graphics,graphicx,latexsym,mathrsfs,revsymb,theorem,url,verbatim,epstopdf,cleveref}
\hypersetup{colorlinks,linkcolor={blue},citecolor={blue},urlcolor={red}}

\newenvironment{proof}{\noindent \textbf{{Proof.~} }}

\newtheorem{definition}{Definition}
\newtheorem{proposition}[definition]{Proposition}
\newtheorem{lemma}[definition]{Lemma}

\newtheorem{theorem}[definition]{Theorem}
\newtheorem{corollary}[definition]{Corollary}
\newtheorem{conjecture}[definition]{Conjecture}

\newtheorem{remark}[definition]{Remark}
\newtheorem{example}[definition]{Example}
\newtheorem{question}[definition]{Question}

\def\bcj{\begin{conjecture}}
\def\ecj{\end{conjecture}}
\def\bcr{\begin{corollary}}
\def\ecr{\end{corollary}}
\def\bd{\begin{definition}}
\def\ed{\end{definition}}
\def\bea{\begin{eqnarray}}
\def\eea{\end{eqnarray}}
\def\bem{\begin{enumerate}}
\def\eem{\end{enumerate}}
\def\bex{\begin{example}}
\def\eex{\end{example}}
\def\bim{\begin{itemize}}
\def\eim{\end{itemize}}
\def\bl{\begin{lemma}}
\def\el{\end{lemma}}
\def\bpf{\begin{proof}}
\def\epf{\end{proof}}
\def\bpp{\begin{proposition}}
\def\epp{\end{proposition}}
\def\bqu{\begin{question}}
\def\equ{\end{question}}
\def\br{\begin{remark}}
\def\er{\end{remark}}
\def\bt{\begin{theorem}}
\def\et{\end{theorem}}
\def\btb{\begin{tabular}}
\def\etb{\end{tabular}}

\def\tr{\mathop{\rm Tr}}

\begin{document}

\title{Detection of genuine multipartite entanglement based on local sum uncertainty relations}

\author{Jun Li}
\thanks{Corresponding author: junlimath@buaa.edu.cn}
\affiliation{School of Mathematical Sciences, Beihang University, Beijing 100191, China}
\author{Lin Chen}
\thanks{Corresponding author: linchen@buaa.edu.cn}
\affiliation{School of Mathematical Sciences, Beihang University, Beijing 100191, China}
\affiliation{International Research Institute for Multidisciplinary Science, Beihang University, Beijing 100191, China}

\begin{abstract}
Genuine multipartite entanglement (GME) offers more significant advantages in quantum information compared with entanglement. We propose a sufficient criterion for the detection of GME based on local sum uncertainty relations for chosen observables of subsystems. We apply the criterion to detect the GME properties of noisy $n$-partite W state when $n = 3, 4, 5$ and $6$, and find that the criterion can detect more noisy W states when $n$ ranges from 4 to 6. Moreover, the criterion is also used to detect the genuine entanglement of $3$-qutrit state. The result is stronger than that based on GME concurrence and fisher information.
\end{abstract}
\maketitle

\section{INTRODUCTION}
\label{sec:intro}

Quantum entanglement \cite{MAN,RPMK,FMA,KSS} is a remarkable resource in the theory of quantum information, which is one of the most distinctive features of quantum theory as compared to classical theory. Entangled states play the essential roles in quantum cryptography \cite{AKE}, teleportation \cite{chbg} and dense coding \cite{chbs}. Genuine multipartite entanglement has more significant advantages compared with entanglement.  It is beneficial in various quantum communication protocols, such as secret sharing \cite{nggr}, extreme spin squeezing \cite{assk}, quantum computing using cluster states \cite{rrhj}, high sensitivity in general metrology tasks \cite{phwl}, and multiparty quantum network \cite{mhvb,vsng}. To certify GME, Bell-like inequalities \cite{jdbn}, various entanglement witness \cite{MF1,JI2,JY3,JS4,CKM5,CEJ6}, and generalized concurrence for multi genuine entanglement \cite{ZH11,YT12,TF13,LM14,LM15} were derived. Some entanglement criteria for bipartite entangled state and multipartite non fully separable states have been also proposed \cite{HPBO,JIV,CYSG}. In particular, the entanglement criteria based on local sum uncertainty relations (LUR) have been proposed for bipartite systems \cite{CJZ} and tripartite systems \cite{YAMA}. Although non-fully separable states contain genuinely entangled states, the criterion of GME based on LUR has not been studied.

In this paper, we study the criterion of GME based on LUR and obtain the sufficient conditions in Theorems \ref{Theorem1} and \ref{Theorem2}.  First, for any quantum states, we show that we can always find the lower bound of LUR. Second, we apply the sum of local observables to the multipartite biseparable state, and obtain the lower bound of LUR by using the method in \cite{CJZ,YAMA}. The converse negative process is  the criterion of detecting GME. When we choose spin observables, the criterion is better than that in \cite{RYMD}. Third, we use the criterion detect $3$-qutrit state and noisy W state for $n$-qubit system ($n=3, 4, 5, 6$) and find that the criterion is strong than the exiting ones \cite{LM14,QIP2020}. Moreover, it can detect more noisy W states when $n$ changes from 4 to 6, which is consistent with the fully-separability of noisy W states \cite{xyc2020}.

In the rest of this paper, we will introduce the criterion of bipartite separability and tripartite fully separability based on LUR in Sec. \ref{sec:perli}. We investigate the GME criterion based on LUR in Sec. \ref{sec:result}, that is, Theorem \ref{Theorem1} and Theorem \ref{Theorem2}. In Sec. \ref{sec:app}, we apply the criterion to noisy W state and $3$-qutrit state to verify its effectiveness. We conclude in Sec. \ref{sec:conclusion}.

\section{PRELIMINARIES}
\label{sec:perli}

 A multipartite state that is not the convex sum of bipartite product states is said to be genuine multipartite entangled \cite{OGGT}. Take the tripartite system as an example. Let $H^d_A$, $H^d_B$, $H^d_C$ denote $\emph{d}-$ dimensional Hilbert spaces of system $A$, $B$, $C$, respectively. A tripartite state $\rho \in \mathfrak{B}(H^d_A\otimes H^d_B\otimes H^d_C)$ is biseparable if it can be expressed
\begin{eqnarray}\label{mo0}
\rho_{BS}=P_{1}\sum_R \eta_{R}^{(1)}\rho_{1}^{R}\otimes\rho_{23}^{R}+P_{2}\sum_{R'} \eta_{R'}^{(2)}\rho_{2}^{R'}\otimes\rho_{13}^{R'}+P_{3}\sum_{R''} \eta_{R''}^{(3)}\rho_{3}^{R''}\otimes\rho_{12}^{R''},
\end{eqnarray}
where $\sum_{k=1}^{3}P_{k}=1, P_{k}\geq 0,$ and $\sum_R \eta_{R}^{(k)}=1.$ Here $\rho_{mn}^{R}$ is an arbitrary density operator for the subsystems $m$ and $n$. Otherwise, $\rho$ is called genuinely tripartite entangled. The definition can be extended to genuine multipartite entangled states. Next, we introduce the criterion of bipartite separability and tripartite fully separability based on local sum uncertainty relations. They can also be detected for the criterion of GME.

Consider the set of local observables $\{A_k\}$ and $\{B_k\}$ for systems $A$ and $B$ respectively. The sum uncertainty relations for arbitrary state $\rho$ are as follows
\begin{eqnarray}\label{mo1}
\sum_k\Delta A^2_k\geq U_A,         \quad \quad \quad      \sum_k\Delta B^2_k\geq U_B ,
\end{eqnarray}
where the non-negative constants $U_A$ and $U_B$ are independent of $\rho$ and $\Delta O^2_k=\langle O^2_k\rangle-\langle O_k\rangle^2=\tr(O^2_k\rho)-\tr^{2}(O_k\rho)$ with $O\in\{A,B\}$. An entanglement criterion based on local sum uncertainty relation was introduced for bipartite system $AB$.

\bl
\label{Lemma1}
\cite{CJZ} For bipartite separable state $\rho_{AB}$, the following inequality holds,
\begin{eqnarray}\label{mo2}
F^{AB}_{\rho_{AB}}
&&:=\sum_k\Delta (A_k\otimes I+I\otimes B_k)^2-(U_A+U_B+M^2_{AB})\nonumber\\
&&\geq 0,
\end{eqnarray}
where $M_{AB}=\sqrt{\sum_k\Delta A^2_k-U_A}-\sqrt{\sum_k\Delta B^2_k-U_B}$. The violation of inequality implies entanglement of $\rho_{AB}$.
\el

For tripartite system, we consider the set of local observables $\{A_k\}$, $\{B_k\}$ and $\{C_k\}$ for subsystem $A$, $B$ and $C$, respectively. Suppose that the sum uncertainty relations for these observables have non-negative constants bounds $U_A$, $U_B$ and $U_C$ independent of states, i. e.
\begin{eqnarray}\label{mo3}
\sum_k\Delta A^2_k\geq U_A,      \quad         \sum_k\Delta B^2_k\geq U_B ,              \quad       \sum_k\Delta C^2_k\geq U_C.
\end{eqnarray}

Recently, the criterion (\ref{mo2}) has been extended to a non-fully separability criterion for the tripartite system based on local sum uncertainty relations as follows.

\bl
\label{Lemma2}
\cite{YAMA} For any tripartite fully separable state $\rho_{ABC}$,
\begin{eqnarray}\label{mo4}
\rho_{ABC}=\sum_ip_i\rho^A_i\otimes \rho^B_i\otimes \rho^C_i,
\end{eqnarray}
the reduced states $\rho_{AB}$, $\rho_{AC}$ and $\rho_{BC}$ are also separable. Therefore, $\rho_{AB}$ must satisfy the inequality (\ref{mo2}) and also similar statements must hold for $\rho_{AC}$ and $\rho_{BC}$. That is,
\begin{eqnarray}\label{mo5}
F_{\rho_{AB}}^{AB}\geq 0,  \quad  F_{\rho_{AC}}^{AC}\geq 0, \quad  F_{\rho_{BC}}^{BC}\geq 0,
\end{eqnarray}
where $F_{\rho_{AC}}^{AC}$ and $F_{\rho_{BC}}^{BC}$ have similar definitions with $F_{\rho_{AB}}^{AB}$.
So the following inequalities must hold simultaneously,
\begin{eqnarray}\label{mo6}
F_{\rho_{ABC}}^{AB|C}\geq 0,  \quad  F_{\rho_{ABC}}^{AC|B}\geq 0,  \quad  F_{\rho_{ABC}}^{BC|A}\geq 0,
\end{eqnarray}
with
$$F_{\rho_{ABC}}^{AB|C}=F_{\rho_{ABC}}-(U_A+U_B+U_C+M^2_{AB}+M^2_{ABC}),$$
$$F_{\rho_{ABC}}^{AC|B}=F_{\rho_{ABC}}-(U_A+U_B+U_C+M^2_{AC}+M^2_{ACB}),$$
$$F_{\rho_{ABC}}^{BC|A}=F_{\rho_{ABC}}-(U_A+U_B+U_C+M^2_{BC}+M^2_{BCA}),$$
where
\begin{eqnarray}\label{ff}
F_{\rho_{ABC}}=\sum_k\Delta (A_k\otimes I_{BC}+B_k\otimes I_{AC}+I_{AB}\otimes C_k)^2_\rho,
\end{eqnarray}
and
$$M_{ABC}=\sqrt{F_{\rho_{AB}}^{AB}}-\sqrt{\sum_k\Delta C^2_k-U_C},$$
and $M_{ACB}$ and $M_{BCA}$ have similar definitions. Violation of any inequality in Eqs. (\ref{mo5}) and (\ref{mo6}) implies non fully separability of $\rho_{ABC}$.
\el

The method of Lemma \ref{Lemma1} and \ref{Lemma2} can be used to find the criterion of genuine entanglement in Theorem \ref{Theorem1}. It may be related to the lower bounds of quantum uncertainty relations for single system and bipartite system. In Eqs. (\ref{mo1}) and (\ref{mo3}), the lower bound of uncertainty relations $U_A$, $U_B$ and $U_C$ are also independent of states. Moreover, some lower bound related to states have also been studied. We know some well-known formula of uncertainty relation for two observables \cite{MLPA},
\begin{eqnarray*}
&&(\Delta A)^{2}+(\Delta B)^{2}\geq \pm i\langle\psi|[A,B]|\psi\rangle+|\langle\psi|A+iB|\psi^{\bot}\rangle|^{2},\nonumber\\
&&(\Delta A)^{2}+(\Delta B)^{2}\geq \frac{1}{2}|\langle\psi_{A+B}^{\bot}|A+B|\psi\rangle|^{2}=\frac{1}{2}[\Delta(A+B)]^{2},
\end{eqnarray*}
where $\langle\psi|\psi^{\bot}\rangle=0$, $|\psi_{A+B}^{\bot}\rangle\varpropto (A-B-\langle A+B\rangle)|\psi\rangle$, and the sign on the right hand side of the inequality takes $+(-)$ while $i[A,B]$ is positive (negative). Let us mark the right side of the inequality as $U_{\rho}$. Furthermore, some multiple observables uncertainty relations were proposed \cite{BCNP,BCSM,QCSL}. We consider the local observables $\{A_{k}\}$ and $\{B_{k}\}$ for systems $A$ and $B$ respectively, the multi-observables sum uncertainty relations are as follows
\begin{eqnarray}\label{mo7}
\sum_k\Delta A^2_k\geq U_{\rho_{A}},       \quad \quad        \sum_k\Delta B^2_k\geq U_{\rho_{B}}.
\end{eqnarray}

Similarly, for bipartite states, the multi-observables sum uncertainty relations are as follows
\begin{eqnarray}\label{mo8}
\sum_k\Delta (A_k\otimes I+I\otimes B_k)^2\geq U_{\rho_{AB}},
\end{eqnarray}
where $U_{\rho_{A}}$, $U_{\rho_{B}}$, and $U_{\rho_{AB}}$ can be obtained by the right side of multi-observables sum uncertainty relations in \cite{BCNP,BCSM,QCSL}.  We will use the forementioned notions and facts in the next section.

\section{MAIN RESULTS}
\label{sec:result}

In this section, we investigate the genuine tripartite and multipartite entanglement based on local sum uncertainty relations. We apply the observables in Eq. (\ref{ff}) to the tripartite biseparable state, and obtain the lower bound of inequality by using Eqs. (\ref{mo7}) and (\ref{mo8}). Thus, we construct a sufficient condition for genuine tripartite entanglement in Theorem \ref{Theorem1}. Further, we extend this criterion to multipartite system in Theorem \ref{Theorem2}.

\subsection{CRITERIA FOR GENUINE TRIPARTITE ENTANGLEMENT}

\bt
\label{Theorem1}
For a tripartite quantum state $\rho_{ABC}$, let Eqs. (\ref{mo7}) and (\ref{mo8}) be satisfied. If $\rho_{ABC}$ is biseparable, then
\begin{eqnarray}\label{Th}
F_{\rho_{ABC}}\geq \min&\{U_{\rho_{A}}+U_{\rho_{BC}}+W^2_{ABC}, \nonumber\\
&U_{\rho_{B}}+U_{\rho_{AC}}+W^2_{BAC},\nonumber\\
&U_{\rho_{C}}+U_{\rho_{AB}}+W^2_{CAB}\}
\end{eqnarray}
where $F_{\rho_{ABC}}$ is defined in (\ref{ff}), and
\begin{eqnarray*}
&&W_{ABC}=\sqrt{\sum_k\Delta (A_k)^2_{\rho_A}-U_{\rho_{A}}} \\
&&-\sqrt{\sum_k\Delta (B_k\otimes I_C+I_B\otimes C_k)^2_{\rho_{BC}}-U_{\rho_{BC}}}, \\
\end{eqnarray*}
and $W_{BAC}$ and $W_{CAB}$ can be similarly defined.
\et

\bpf
If $\rho_{ABC}$ is biseparable, it can be written as Eq. (\ref{mo0}) \cite{JDB,BJTM,RYMD},
\begin{eqnarray*}
\rho_{BS}=&&P_{1}\sum_R \eta_{R}^{1}\rho_{1}^{R}\otimes\rho_{23}^{R}+P_{2}\sum_{R'} \eta_{R'}^{2}\rho_{2}^{R'}\otimes\rho_{13}^{R'}\nonumber\\
&&+P_{3}\sum_{R''} \eta_{R''}^{3}\rho_{3}^{R''}\otimes\rho_{12}^{R''}
\end{eqnarray*}
with $0\leq P_k\leq 1$ , $\sum_k P_k =1$ and $\sum_R \eta_{R}^{k}=1$.

For any mixture of type $\rho_{mix}=\sum_{R\geq 1} P_{R}\rho^{R}$, the variance $\Delta^{2}u$ satisfies \cite{HFHS}
\begin{eqnarray}\label{mo10}
\Delta^{2}u \geq \sum_R P_{R} \Delta^{2}_{u_{R}}.
\end{eqnarray}

Hence for the biseparable state,
\begin{eqnarray}\label{mo11}
&&\sum_k\Delta (A_k\otimes I_{BC}+B_k\otimes I_{AC}+I_{AB}\otimes C_k)^2_{\rho_{BS}}\nonumber\\
&&\geq P_{1} \sum_k\Delta (A_k\otimes I_{BC}+B_k\otimes I_{AC}+I_{AB}\otimes C_k)^2_{\rho_{R}}\nonumber\\
&&+ P_{2} \sum_k\Delta (A_k\otimes I_{BC}+B_k\otimes I_{AC}+I_{AB}\otimes C_k)^2_{\rho_{R^{'}}}\nonumber\\
&&+ P_{3} \sum_k\Delta (A_k\otimes I_{BC}+B_k\otimes I_{AC}+I_{AB}\otimes C_k)^2_{\rho_{R^{''}}}\nonumber\\
&&\geq \min\{\sum_k\Delta (A_k\otimes I_{BC}+B_k\otimes I_{AC}+I_{AB}\otimes C_k)^2_{\rho_{R}},\nonumber\\
&&\quad \quad \quad\sum_k\Delta (A_k\otimes I_{BC}+B_k\otimes I_{AC}+I_{AB}\otimes C_k)^2_{\rho_{R^{'}}},\nonumber\\
&&\quad \quad \quad\sum_k\Delta (A_k\otimes I_{BC}+B_k\otimes I_{AC}+I_{AB}\otimes C_k)^2_{\rho_{R^{''}}}\}.
\end{eqnarray}

We can always choose as the lower bound the smallest value of $\sum_k\Delta (A_k\otimes I_{BC}+B_k\otimes I_{AC}+I_{AB}\otimes C_k)^2_{\rho_{\zeta}}$ in (\ref{mo11}). So the second inequality can be obtained using the fact that $\sum_k P_k =1$.

Then we consider $\sum_k\Delta (A_k\otimes I_{BC}+B_k\otimes I_{AC}+I_{AB}\otimes C_k)^2$ that corresponds to the bipartition $\sum_R\eta_{R}^{1}\rho_{1}^{R}\otimes\rho_{23}^{R}$,
\begin{eqnarray}\label{mo12}
&&\sum_k\Delta (A_k\otimes I_{BC}+B_k\otimes I_{AC}+I_{AB}\otimes C_k)^2_{\rho_{R}}\nonumber\\
&&=\sum_k\{\langle[A_k\otimes I_{BC}+I_A\otimes (B_k\otimes I_C+I_B\otimes C_k)]^2\rangle\nonumber\\
&&-\langle A_k\otimes I_{BC}+I_A\otimes (B_k\otimes I_C+I_B\otimes C_k)\rangle^2\}\nonumber\\
&&=\sum_k\Delta (A_k)^2_{\rho_{A}}+\sum_k\Delta (B_k\otimes I_C+I_B\otimes C_k)^2_{\rho_{BC}}\nonumber\\
&&+2\sum_k[\langle A_k\otimes (B_k\otimes I_C+I_B\otimes C_k)\rangle-\nonumber\\
&&\langle A_k\otimes I_{BC}\rangle\langle I_A\otimes (B_k\otimes I_C+I_B\otimes C_k)\rangle]\nonumber\\
&&\geq \sum_k\Delta (A_k)^2_{\rho_{A}}+\sum_k\Delta (B_k\otimes I_C+I_B\otimes C_k)^2_{\rho_{BC}}\nonumber\\
&&-2\sqrt{[\sum_k\Delta (A_k)^2_{\rho_{A}}-U_{\rho_{A}}]}\cdot\sqrt{[\sum_k\Delta (B_k\otimes I_C+I_B\otimes C_k)^2_{\rho_{BC}}-U_{\rho_{BC}}]}\nonumber\\
&&=U_{\rho_{A}}+U_{\rho_{BC}}+W^2_{ABC},
\end{eqnarray}
where $W_{ABC}=\sqrt{\sum_k\Delta (A_k)^2_{\rho_A}-U_{\rho_{A}}}-\sqrt{\sum_k\Delta (B_k\otimes I_C+I_B\otimes C_k)^2_{\rho_{BC}}-U_{\rho_{BC}}}$. The inequality is due to Lemma 1 in \cite{CJZ}.

Combining Eq. (\ref{mo11}) and Eq. (\ref{mo12}), we can obtain Eq. (\ref{Th}). In Eq. (\ref{Th}), the first term in the bracket $\{\}$, namely, $U_{\rho_{A}}+U_{\rho_{BC}}+W^2_{ABC}$ is implied by the biseparable state $\sum_R\eta_{R}^{1}\rho_{1}^{R}\otimes\rho_{23}^{R}$. Similarly, the second term is implied by the biseparable state $\sum_R\eta_{R'}^{2}\rho_{2}^{R'}\otimes\rho_{13}^{R'}$, and the third term is implied by the biseparable state $\sum_R\eta_{R''}^{3}\rho_{3}^{R''}\otimes\rho_{12}^{R''}$. Violation of the inequality (\ref{Th}) is sufficient to confirm genuine tripartite entanglement of $\rho_{ABC}$.      \quad \quad \quad  $\square$
\epf

When we choose spin observables as the observables $A$, $B$, and $C$, the criteria in Theorem \ref{Theorem1} require only the statistics of a set of observables. In this sense, it is state independent, which is similar to \cite{RYMD}. In order to compare Theorem \ref{Theorem1} with criterion 1 in \cite{RYMD}, we consider the sum of $\Delta^{2}u$ and $\Delta^{2}v$ where
\begin{eqnarray}\label{mo13}
&u=h_{1}J_{x,1}+h_{2}J_{x,2}+h_{3}J_{x,3}\nonumber\\
&v=g_{1}J_{y,1}+g_{2}J_{y,2}+g_{3}J_{y,3}
\end{eqnarray}
and $h_{k}$ and $g_{k}$ $({k=1,2,3})$ are real numbers. Here $J_{x,k}$, $J_{y,k}$, $J_{z,k}$ are the spin operators for subsystem $k$, satisfying $[J_{x,k},J_{y,k}]=J_{z,k}$. Then $F_{\rho_{ABC}}$ in Eq. (\ref{Th}) is equal to $\Delta^{2}u+\Delta^{2}v$ when $A_{1}=h_{1}J_{x,1}$, $B_{1}=h_{2}J_{x,2}$, $C_{1}=h_{3}J_{x,3}$, $A_{2}=g_{1}J_{y,1}$, $B_{2}=g_{2}J_{y,2}$, $C_{2}=g_{3}J_{y,3}$ and $k=2$. This leads us to the following criterion,

\bcr
Violation of the inequality
\begin{eqnarray}\label{mo14}
\Delta^{2}u+\Delta^{2}v\geq \min&&\{|g_{1}h_{1}\langle J_{z,1}\rangle|+|g_{2}h_{2}\langle J_{z,2}\rangle+g_{3}h_{3}\langle J_{z,3}\rangle|+W^{2}_{123},\nonumber\\
&&|g_{2}h_{2}\langle J_{z,2}\rangle|+|g_{1}h_{1}\langle J_{z,1}\rangle+g_{3}h_{3}\langle J_{z,3}\rangle|+W^{2}_{213},\nonumber\\
&&|g_{3}h_{3}\langle J_{z,3}\rangle|+|g_{1}h_{1}\langle J_{z,1}\rangle+g_{2}h_{2}\langle J_{z,2}\rangle|+W^{2}_{312}\}
\end{eqnarray}
is sufficient to confirm genuine tripartite entanglement. Here
\begin{eqnarray*}
&&W_{123}=\sqrt{\Delta^{2}(h_{1}J_{x,1})+\Delta^{2}(g_{1}J_{y,1})-|g_{1}h_{1}\langle J_{z,1}\rangle|}-\nonumber\\
&&\sqrt{\Delta^{2}(h_{2}J_{x,2}+h_{3}J_{x,3})+\Delta^{2}(g_{2}J_{y,2}+g_{3}J_{y,3})-|g_{2}h_{2}\langle J_{z,2}\rangle+g_{3}h_{3}\langle J_{z,3}\rangle|},
\end{eqnarray*}
$W_{213}$ and $W_{312}$ can be similarly defined.
\ecr

If $W_{123}=W_{213}=W_{312}=0$, Eq. (\ref{mo14}) is reduced to the result in \cite{RYMD}, so the above criterion is better than criterion 1 in \cite{RYMD}. For the specific spin state, we can choose the optimal values for $h_{k}$, $g_{k}$.

\subsection{CRITERIA FOR GENUINE MULTIPARTITE ENTANGLEMENT}

Now we extend the method in previous section used to derive criteria for genuine tripartite entanglement to $N$-partite system. One can show that the number of possible bipartition is $2^{N-1}-1$. In order to investigate the criteria of genuine $N$-partite entanglement, we should consider every bipartition. Here we generalize the criterion in Eq. (\ref{Th}) and Eq. (\ref{mo14}) for $N$-partite system. We denote every bipartition by $S_{r}-S_{s}$, where $S_{r}$ and $S_{s}$ are the sets of two part in a specific bipartition.

\bt
\label{Theorem2}
If a $N$-partite quantum state $\rho_{A_{1}A_{2}\ldots A_{N}}$ is biseparable, then
\begin{eqnarray}\label{mo15}
F_{\rho_{A_{1}A_{2}\ldots A_{N}}}\geq \min \{U_{BS}\},
\end{eqnarray}
where $U_{BS}$ is the set of the quantity $U_{\rho_{k_{r}}}+U_{\rho_{k_{s}}}+W^{2}_{\rho_{k_{r}|k_{s}}}$ defined for each partition $S_{r}-S_{s}$, $\rho_{k_{r}}$ and $\rho_{k_{s}}$ are the states in set $S_{r}$ and $S_{s}$ respectively. Violation of the inequality (\ref{mo15}) is sufficient to confirm genuine $N$-partite entanglement. The proof of the inequality follows from the proof in Eq. (\ref{Th}).
\et

When the observables in Eq. (\ref{mo15}) are spin observables, the following criterion can be obtained.

\bcr
Violation of the inequality
\begin{eqnarray}\label{mo16}
\Delta^{2}u+\Delta^{2}v\geq \min\{S_{B}\}
\end{eqnarray}
implies genuine $N$-partite entanglement. Where $S_{B}$ is the set of $|\Sigma_{k_{r}=1}^{m}h_{k_{r}}g_{k_{r}}\langle J_{z,k_{r}}\rangle|+|\Sigma_{k_{s}=1}^{n}h_{k_{s}}g_{k_{s}}\langle J_{z,k_{s}}\rangle|+W^{2}_{\rho_{k_{r}|k_{s}}}$ defined for each partition $S_{r}-S_{s}$. When every $W_{\rho_{k_{r}|k_{s}}}=0$, the inequality is reduced to criterion 6 in \cite{RYMD}.
\ecr

For $N=4$, there will be $2^{4-1}-1=7$ bipartition. They are $1-234$, $2-134$, $3-124$, $4-123$, $12-34$, $13-24$, $14-23$. Using them we obtain the criterion for genuine four-partite entanglement.

\bcr
If a four-partite quantum state is biseparable, then
\begin{eqnarray}\label{mo17}
\Delta^{2}u+\Delta^{2}v\geq \min&&\{|g_{1}h_{1}\langle J_{z,1}\rangle|+|g_{2}h_{2}\langle J_{z,2}\rangle+g_{3}h_{3}\langle J_{z,3}\rangle+g_{4}h_{4}\langle J_{z,4}\rangle|+W^{2}_{1|234},\nonumber\\
&&|g_{2}h_{2}\langle J_{z,2}\rangle|+|g_{1}h_{1}\langle J_{z,1}\rangle+g_{3}h_{3}\langle J_{z,3}\rangle+g_{4}h_{4}\langle J_{z,4}\rangle|+W^{2}_{2|134},\nonumber\\
&&|g_{3}h_{3}\langle J_{z,3}\rangle|+|g_{1}h_{1}\langle J_{z,1}\rangle+g_{2}h_{2}\langle J_{z,2}\rangle+g_{4}h_{4}\langle J_{z,4}\rangle|+W^{2}_{3|124},\nonumber\\
&&|g_{4}h_{4}\langle J_{z,4}\rangle|+|g_{1}h_{1}\langle J_{z,1}\rangle+g_{2}h_{2}\langle J_{z,2}\rangle+g_{3}h_{3}\langle J_{z,3}\rangle|+W^{2}_{4|123},\nonumber\\
&&|g_{1}h_{1}\langle J_{z,1}\rangle+g_{2}h_{2}\langle J_{z,2}\rangle|+|g_{3}h_{3}\langle J_{z,3}\rangle+g_{4}h_{4}\langle J_{z,4}\rangle|+W^{2}_{12|34},\nonumber\\
&&|g_{1}h_{1}\langle J_{z,1}\rangle+g_{3}h_{3}\langle J_{z,3}\rangle|+|g_{2}h_{2}\langle J_{z,2}\rangle+g_{4}h_{4}\langle J_{z,4}\rangle|+W^{2}_{13|24},\nonumber\\
&&|g_{1}h_{1}\langle J_{z,1}\rangle+g_{4}h_{4}\langle J_{z,4}\rangle|+|g_{2}h_{2}\langle J_{z,2}\rangle+g_{3}h_{3}\langle J_{z,3}\rangle|+W^{2}_{14|23}\}
\end{eqnarray}
where
\begin{eqnarray*}
&&W_{1|234}=\sqrt{\Delta^{2}(h_{1}J_{x,1})+\Delta^{2}(g_{1}J_{y,1})-|g_{1}h_{1}\langle J_{z,1}\rangle|}-\nonumber\\
&&\sqrt{\Delta^{2}(h_{2}J_{x,2}+h_{3}J_{x,3}+h_{4}J_{x,4})+\Delta^{2}(g_{2}J_{y,2}+g_{3}J_{y,3}+g_{4}J_{y,4})-|g_{2}h_{2}\langle J_{z,2}\rangle+g_{3}h_{3}\langle J_{z,3}\rangle+g_{4}h_{4}\langle J_{z,4}\rangle|},
\end{eqnarray*}
$W_{2|134}$, $W_{3|124}$, $W_{4|123}$, $W_{12|34}$, $W_{13|24}$, and $W_{14|23}$ can be  similarly defined.
 \ecr

The violation of the inequality in Eq. (\ref{mo17}) implies genuine four-partite entanglement. If $W_{1|234}=W_{2|134}=W_{3|124}=W_{4|123}=W_{12|34}=W_{13|24}=W_{14|23}=0$, Eq. (\ref{mo17}) is reduced to the criterion 8 in \cite{RYMD}, so the above inequality is better than the result in \cite{RYMD}.

\section{EXAMPLE}
\label{sec:app}

In this section, we illustrate the utility of the criteria by a few examples.

\bex
Consider the $n$-qubit W state mixed with the white noise,
$$\rho_{W_{n}}(q)=\frac{1-q}{2^{n}}I+q|W_{n}\rangle\langle W_{n}|,$$
where $0\leq q\leq 1$, $|W_{n}\rangle=\frac{1}{\sqrt{2}}(|10\ldots00\rangle+|01\ldots00\rangle+\ldots+|00\ldots01\rangle) $ and $I$ is the $2^{n} \times 2^{n}$ identity matrix.
\eex

\begin{figure}[h]
\centering
\includegraphics[width=10in]{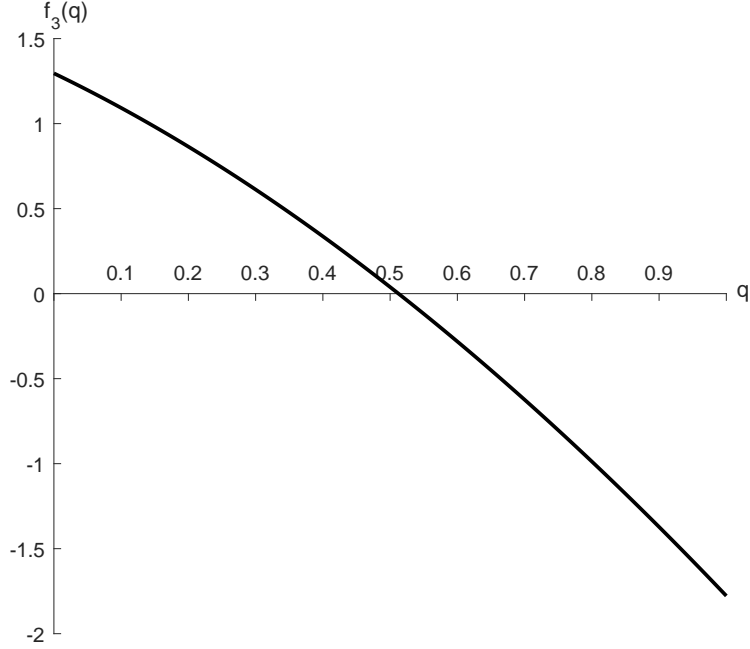}
\caption{The abscissa and ordinate represent $q$ and $f(q)$, respectively. Below the abscissa axis means that Theorem \ref{Theorem1} can detect genuinely entangled state for $0.512\leq q\leq 1$.}
\label{fig:WS}
\end{figure}

Set $A_{1}=B_{1}=-C_{1}=\sigma_{x}$, $A_{2}=B_{2}=-C_{2}=\sigma_{y}$, and $A_{3}=B_{3}=C_{3}=\sigma_{z}$. The criterion in Theorem \ref{Theorem1} is computed to be
$f(q)=-q^{2}-\frac{14}{3}q+\frac{35}{9}-(\sqrt{\frac{1}{9}-\frac{1}{9}q^{2}}-\sqrt{-\frac{4}{9}q^{2}-\frac{10}{3}q+\frac{34}{9}})^{2}$, which means the left side of (\ref{Th}) minus the right of that, as shown in FIG \ref{fig:WS}. Comparing with the criterion in \cite{LM14} and \cite{QIP2020}, Theorem \ref{Theorem1} can detect more genuinely tripartite entangled states.

Now we consider genuine entanglement of $\rho_{W_{n}}(q)$ for three cases.
\begin{enumerate}
  \item When $n=4$, we set $A_{1}=B_{1}=C_{1}=-D_{1}=\sigma_{x}$, $A_{2}=B_{2}=C_{2}=-D_{2}=\sigma_{y}$, and $A_{3}=B_{3}=C_{3}=D_{3}=\sigma_{z}$ in Theorem \ref{Theorem2}. By calculation, we can obtain $f_{4}(q)=-4q^{2}+3-(\sqrt{-q^{2}+2q}-\sqrt{-q^{2}-2q+3})^{2}$, which means the left side of (\ref{mo15}) minus the right side.
  \item when $n=5$, we set $A_{1}=B_{1}=C_{1}=-D_{1}=-E_{1}=\sigma_{x}$, $A_{2}=B_{2}=C_{2}=-D_{2}=-E_{2}=\sigma_{y}$, and $A_{3}=B_{3}=C_{3}=D_{3}=E_{3}=\sigma_{z}$.  By calculation, we can obtain $f_{5}(q)=-9q^{2}+\frac{4}{5}q+\frac{91}{25}-(\sqrt{-\frac{81}{25}q^{2}-\frac{2}{5}q+\frac{91}{25}}-\sqrt{-\frac{36}{25}q^{2}+2q})^{2}$.
  \item When $n=6$, we set $A_{1}=B_{1}=C_{1}=-D_{1}=-E_{1}=-F_{1}=\sigma_{x}$, $A_{2}=B_{2}=C_{2}=-D_{2}=-E_{2}=-F_{2}=\sigma_{y}$, and $A_{3}=B_{3}=C_{3}=D_{3}=E_{3}=F_{3}=\sigma_{z}$.  By calculation, we can obtain $f_{6}(q)= -16q^{2}+6q+\frac{77}{9}-(\sqrt{-\frac{100}{9}q^{2}+4q+\frac{64}{9}}-\sqrt{\frac{4}{9}-\frac{4}{9}q^{2}})^2$.
\end{enumerate}

\begin{figure}[ht]
\centering
\includegraphics[width=10in]{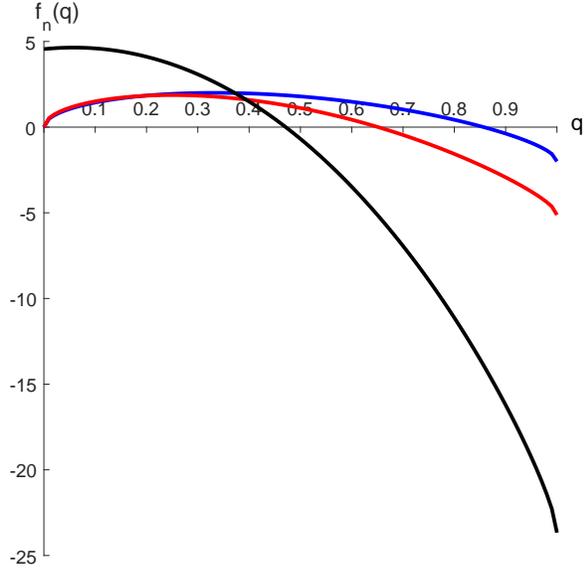}
\caption{The abscissa and ordinate represent $q$ and $f_{n}(q)$, respectively. The blue, red and black lines represent $f_{4}(q)$, $f_{5}(q)$, and $f_{6}(q)$, respectively. Below the abscissa axis means that Theorem \ref{Theorem2} can detect genuinely entangled state for $0.857\leq q\leq 1$ when $n=4$. Similarly, we have $0.651\leq q\leq 1$ when $n=5$, and $0.46\leq q\leq 1$ when $n=6$. With the increase of $n$, more genuinely entangled states can be detected. }
\label{fig:w456}
\end{figure}

We describe the three cases in FIG \ref{fig:w456}. The same method can be used when $n\geq 7$ by choosing appropriate observables, but the calculation will become more and more complex.

\begin{figure}[htb]
\centering
\includegraphics[width=10in]{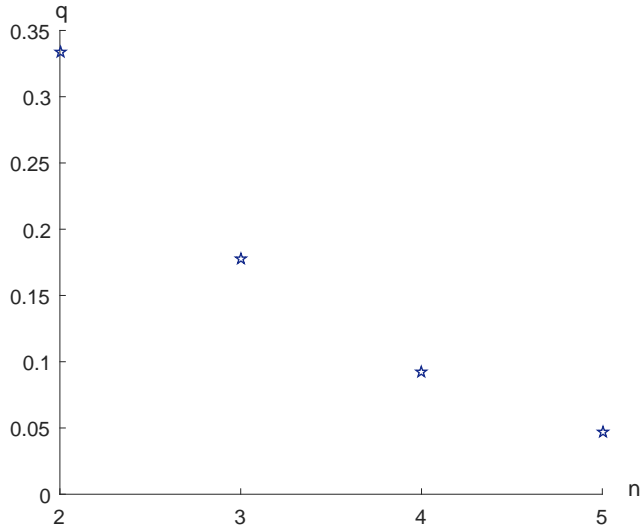}
\caption{The abscissa and ordinate represent critical value $q$ and the number of system$n$, respectively. Above the stars are entangled states that can be detected. With the increase of $n$, the criterion can detect more states.}
\label{fig:wsepn}
\end{figure}

\clearpage

It is worth mentioning that \cite{xyc2020} the noisy $W$ state $\rho_{W_{n}}(q)$ is fully separable if
\begin{eqnarray}\label{ex1}
q\leq \left\{\begin{array}{cc}
              \frac{1}{1+2^{n}\sqrt{\frac{n-1}{2n}}} & if \quad 2\leq n\leq5; \\
              \frac{n}{n+(n-2)2^{n}} & if \quad n\geq 6.
            \end{array} \right.
\end{eqnarray}
The condition is necessary and sufficient when $n\leq 5$. This is similar to the genuine entanglement criterion in Theorems \ref{Theorem1} and \ref{Theorem2}, that is, (\ref{ex1}) can detect more states with the increase of $n$. We describe these results in FIG \ref{fig:wsepn}.

This criterion can not only detect the GME of qubit states, but also detect that of qurit states. Here is an example of the 3-qutrit state.

\bex
Consider a $3$-qutrit state mixed with the white noise \cite{LM14},
$$\rho=\frac{1-x}{27}I+x|\varphi\rangle\langle \varphi|,$$
where $0\leq x\leq 1$, $|\varphi\rangle=\frac{1}{\sqrt{3}}(|012\rangle+|021\rangle+|111\rangle) $ and $I$ is the $3^{3} \times 3^{3}$ identity matrix.
\eex

\begin{figure}[ht]
\centering
\includegraphics[width=10in]{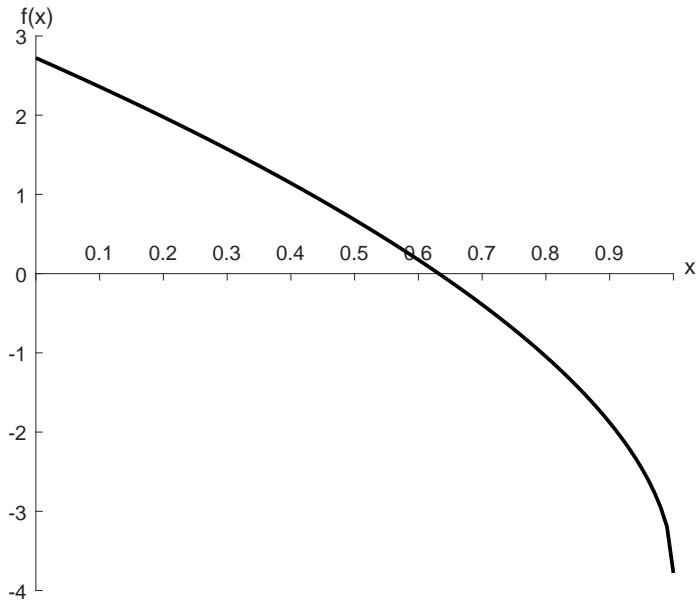}
\caption{The abscissa and ordinate represent $x$ and $f(x)$, respectively. Below the abscissa axis means that Theorem \ref{Theorem1} can detect genuinely entangled state for $0.632\leq x\leq 1$.}
\label{fig:qutrit}
\end{figure}

Set $A_{1}=-B_{1}=-C_{1}=J_{x}$, $A_{2}=-B_{2}=-C_{2}=J_{y}$, and $A_{3}=B_{3}=C_{3}=J_{z}$. Here $J_{x}$, $J_{y}$, $J_{z}$ are spin operators. The criterion in Theorem \ref{Theorem1} is computed to be
$f(x)=\frac{25}{9}-4x-(\sqrt{-\frac{x^{2}}{9}-3x+\frac{28}{9}}-\sqrt{-\frac{x^{2}}{9}+\frac{x}{3}+\frac{7}{3}})^{2}$, which means the left side of (\ref{Th}) minus the right of that, as shown in FIG \ref{fig:qutrit}. The criterion can detect GME better than the criterion in \cite{LM14}.

\section{CONCLUSION}
\label{sec:conclusion}

The detection of GME is a basic and important object in quantum theory. In view of the bipartite entanglement and tripartite non-fully separable criteria based on LUR, we have studied the GME based on LUR. We have obtained an effective criterion to detecting the GME for tripartite system, which be extended to multipartite system. Comparing with some existing criteria, the criterion can detect more genuinely entangled states by theoretical analysis and numerical examples. Also, we found the relation of $n$ and genuinely entanglement for $n$- qubit noisy W state. The method used in this paper can also be generalized to arbitrary multipartite qudit systems. It would be also worthwhile to investigate the $k$-separability of multipartite systems.

\bigskip
\noindent{\bf Acknowledgments}\,  Authors were supported by the NNSF of China (Grant No. 11871089), and the Fundamental
Research Funds for the Central Universities (Grant Nos. KG12080401 and ZG216S1902).

\end{document}